\documentclass[10pt,twocolumn,superscriptaddress,amsmath,amssymb,aps,pra]{revtex4-2}
\usepackage{amsmath}
\usepackage{graphicx}% Include figure files
\usepackage{booktabs} % Required for \midrule in tables
\usepackage{dcolumn}% Align table columns on decimal point
\usepackage{physics}
\usepackage{comment}
\usepackage[T1]{fontenc}
\usepackage{mathtools}
\usepackage[mode = text]{siunitx}
\usepackage{newtx}

\usepackage[dvipsnames]{xcolor}
\definecolor{dark-blue}{rgb}{0,0.2,0.6}
\usepackage[colorlinks,%
            pdfusetitle,%
            urlcolor=dark-blue,%
            citecolor=dark-blue,%
            linkcolor=dark-blue]{hyperref}
\usepackage{orcidlink}

\usepackage{etoolbox}
\makeatletter
\pretocmd{\NAT@open}{\begingroup\color{\@citecolor}}{}{}
\apptocmd{\NAT@close}{\endgroup}{}{}
\makeatother

\newcommand{\nm}{\nano\meter}
\newcommand{\us}{\micro \second}
\newcommand{\um}{\micro\meter}

\newcommand{\uK}{\micro\kelvin}
\newcommand{\mK}{\milli\kelvin}

\begin{document}

\title{Microsecond-scale high-survival and number-resolved detection of ytterbium atom arrays}

\author{A.~\surname{Muzi~Falconi}~\orcidlink{0009-0008-5242-9016}}
\affiliation{Department of Physics, University of Trieste, 34127 Trieste, Italy}
\author{R.~Panza~\orcidlink{0009-0008-2059-4201}}
\affiliation{Department of Physics, University of Trieste, 34127 Trieste, Italy}
\affiliation{Istituto Nazionale di Ottica del Consiglio Nazionale delle Ricerche (CNR-INO), 34149 Trieste, Italy}
\author{S.~Sbernardori}
\affiliation{Department of Physics, University of Trieste, 34127 Trieste, Italy}
\affiliation{Istituto Nazionale di Ottica del Consiglio Nazionale delle Ricerche (CNR-INO), 34149 Trieste, Italy}
\author{R.~Forti~\orcidlink{0009-0009-9463-3397}}
\affiliation{Department of Physics, University of Trieste, 34127 Trieste, Italy}
\affiliation{Elettra Sincrotrone Trieste S.C.p.A., 34149 Trieste, Italy}
\author{R.~Klemt~\orcidlink{0009-0002-3231-447X}}
\affiliation{\,5.~Physikalisches Institut and Center for Integrated Quantum Science and Technology, Universit\"at Stuttgart, 70569 Stuttgart, Germany}
\author{O.~\surname{Abdel~Karim}}
\affiliation{Istituto Nazionale di Ottica del Consiglio Nazionale delle Ricerche (CNR-INO), 34149 Trieste, Italy}
\author{M.~Marinelli~\orcidlink{0000-0002-1981-8182}}
\affiliation{Department of Physics, University of Trieste, 34127 Trieste, Italy}
\affiliation{Istituto Officina dei Materiali del Consiglio Nazionale delle Ricerche (CNR-IOM), 34149 Trieste, Italy}
\author{F.~Scazza~\orcidlink{0000-0001-5527-1068}}
\email[E-mail address: ] {francesco.scazza@units.it}
\affiliation{Department of Physics, University of Trieste, 34127 Trieste, Italy}
\affiliation{Istituto Nazionale di Ottica del Consiglio Nazionale delle Ricerche (CNR-INO), 34149 Trieste, Italy}

%\date{\today}

\begin{abstract}
Scalable atom-based quantum platforms for simulation, computing, and metrology require fast high-fidelity, low-loss imaging of individual atoms.
Standard fluorescence detection methods rely on continuous cooling, limiting the detection range to one atom and imposing long imaging times that constrain the experimental cycle and mid-circuit conditional operations.
Here, we demonstrate fast and low-loss single-atom imaging in optical tweezers without active cooling, enabled by the favorable properties of ytterbium.
Collecting fluorescence over microsecond timescales, we reach single-atom discrimination fidelities above $99.9\%$ and single-shot survival probabilities above $99.5\%$.
Through interleaved recooling pulses, as short as a few hundred microseconds for atoms in magic traps, we perform tens of consecutive detections with constant atom-retention probability per image---an essential step toward fast atom re-use in tweezer-based processors and clocks.
Our scheme does not induce parity projection in multiply-occupied traps, enabling number-resolved single-shot detection of several atoms per site.
This allows us to study the near-deterministic preparation of single atoms in tweezers driven by blue-detuned light-assisted collisions.
Moreover, the near-diffraction-limited spatial resolution of our low-loss imaging enables number-resolved microscopy in dense arrays, opening the way to direct site-occupancy readout in optical lattices for density fluctuation and correlation measurements in quantum simulators. 
\end{abstract}

\maketitle
In recent years, a new generation of cold atom-based machines has emerged, realizing the vision of scalable quantum many-body systems with resolution at the single-particle level. 
These notably include lattice gas microscopes~\cite{Gross_2017,Gross_2021}, tweezer- and lattice-trapped atom arrays~\cite{Browaeys_2020,Ebadi_2021,Scholl_2021,Kaufman_2021,Nelson_2007,Norcia_2024,Gyger_2024}, and free-space matter-wave microscopes~\cite{Asteria_2021,Holten_2022,Brandstetter_2025,Yao_2025,DeJongh_2025,Xiang_2025}, all of which hinge on the ability to detect individual atoms with high fidelity. 
Such capability is also becoming increasingly important in optical atomic clocks~\cite{Ludlow_2015,Madjarov_2019,Norcia_2019,Young_2020,Finkelstein_2024,Shaw_2024}.
Most single-atom imaging schemes rely on collecting fluorescence under continuous laser cooling to ensure atoms remain localized within their trapping sites. 
While this approach yields high fidelities and atom survival, it requires careful control of trapping light shifts and long imaging durations from ten to hundreds of milliseconds~\cite{Gross_2021,Bakr_2009,Sherson_2010,Yamamoto_2016, Covey_2019,Saskin_2019,Ma_2022,Lis_2023,Huie_2023,Norcia_2023,Tao_2024,Norcia_2024}, which limit the experimental repetition rate and mid-circuit measurement capabilities. 
Moreover, cooling light induces pairwise losses through light-assisted collisions (LACs) in the trap \cite{DePue_1999,Schlosser_2002,Fuhrmanek_2012,Sompet_2013}, restricting measurements to resolve only the on-site atom-number parity.
 
A fundamentally different strategy is to operate in a non-equilibrium, cooling-free regime where atoms scatter a burst of photons in a short time~\cite{Bucker_2009,Miranda_2015,Bergschneider_2018,Su_2023,Su_2025,Ma_2023,Scholl_2023,Senoo_2025,Tao_2025,Grun_2024}. 
This enables faster detection and avoids parity projection.
However, existing implementations have either been fully destructive~\cite{Ma_2023,Scholl_2023,Senoo_2025,Tao_2025}, with atoms being lost from the trap upon excessive heating, or have operated with atoms in free flight~\cite{Bucker_2009,Bergschneider_2018,Su_2025}. 
Atoms escaping their trapping site also deteriorate the imaging spatial resolution, as they diffuse under the effect of photon recoils ~\cite{Bergschneider_2018} and deposit fluorescence on a large detector area, leading to an increased spread of the signal. 
Low-loss fluorescence imaging without cooling becomes possible for atoms like ytterbium, where a strong, closed optical transition facilitates efficient photon scattering, while the large atomic mass reduces recoil-induced heating. 
These properties open the door to fast, high-fidelity, and minimally destructive detection in tightly-spaced tweezer and lattice arrays.
Moreover, gaining on-site number-resolution would greatly benefit studies of itinerant many-body systems~\cite{Boll_2016, Koepsell_2020,Hartke_2020,Su_2023,Prichard_2024,Lebrat_2024} and optical clock architectures~\cite{Marti_2018,Goban_2018,Milner_2023}.

Here, we realize microsecond-scale detection of single ytterbium atoms trapped in optical tweezer arrays with fidelities exceeding $99.9\%$ and less than $0.5\%$ losses.
Our results demonstrate the fastest neutral-atom imaging to date combining high fidelity and high survival, matching state-of-the-art performances in alkaline-earth-like atoms~\cite{Yamamoto_2016, Saskin_2019,Covey_2019,Ma_2022,Huie_2023,Lis_2023,Norcia_2023,Tao_2024} and paralleling the speed of cavity-assisted readout without its added complexity~\cite{Deist_2022, Wang2025}.
While the imaging process injects significant motional excitations, we show that the atoms' temperature can be efficiently restored using a recooling pulse---as short as a few hundred microseconds for magic trapping conditions---allowing us to acquire multiple images with uniform loss probability per detection. 
These capabilities are well-suited for atom rearrangement and, in combination with spin shelving, for fast mid-circuit readout in neutral atom-based quantum information processors~\cite{Lis_2023,Scholl_2023,Ma_2023,Norcia_2023,Deist_2022,Graham_2023,Bluvstein_2024}.
Remarkably, our in-trap detection protocol does not induce parity projection and resolves several atoms per trap.
Equipped with multi-atom detection, we investigate the near-deterministic loading dynamics of single ${}^{171}\text{Yb}$ atoms in tweezers, obtained through the combination of gray molasses cooling and repulsive molecular potentials~\cite{Grunzweig_2010,Lester_2015,Brown_2019,Jenkins_2022}. 
Finally, we combine our imaging technique with a likelihood-based reconstruction algorithm to yield high-fidelity detection of site occupations in dense arrays, featuring an inter-site spacing comparable to those employed in lattice gas microscopes.

\clearpage

\begin{figure}[t]
    \centering\includegraphics[width=\columnwidth]{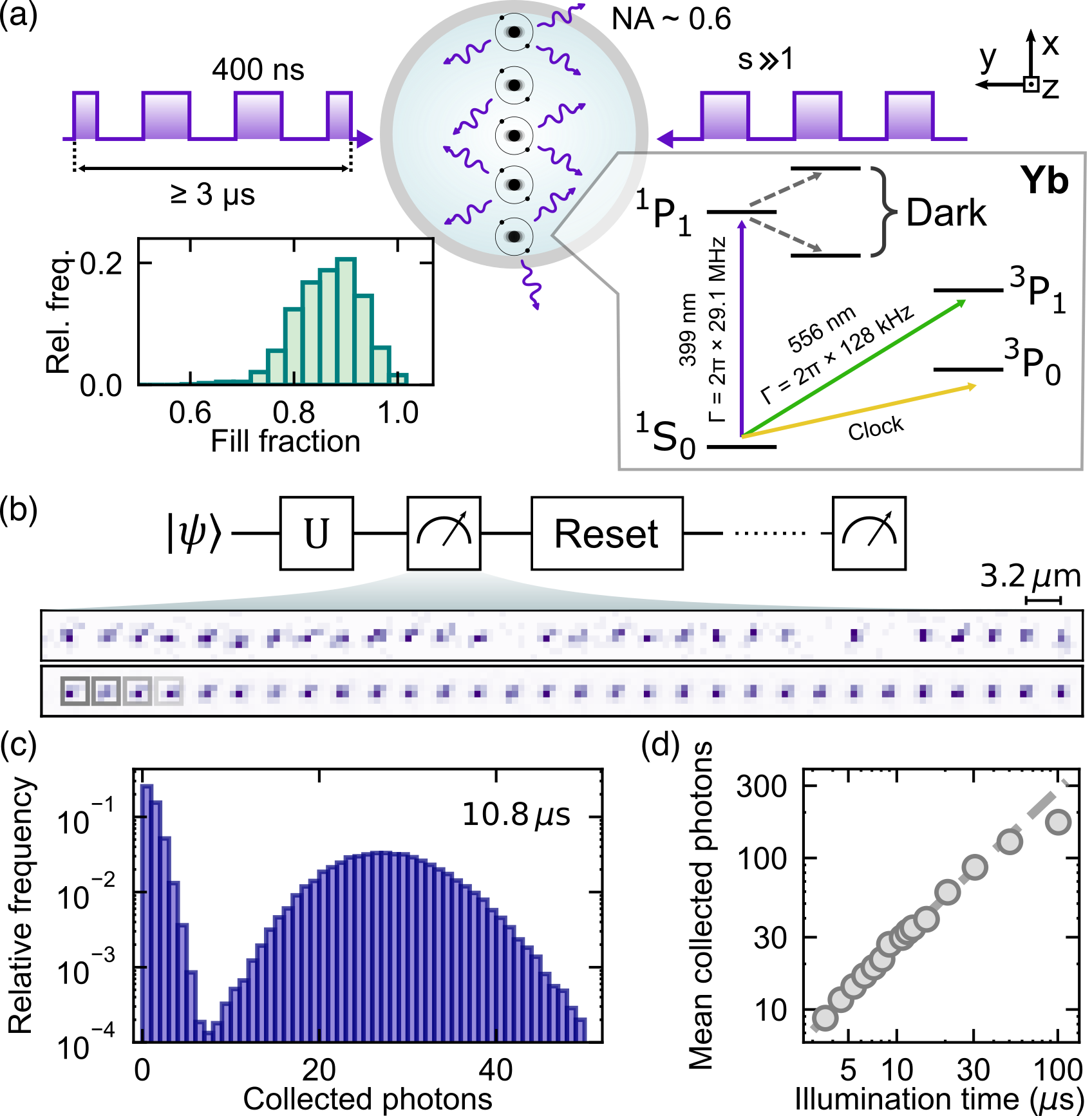}
    \caption{Fast imaging of ytterbium arrays. (a) An array of ytterbium atoms is illuminated with high-intensity ($s\simeq40$) counter-propagating alternated pulses. 
    The scattered photons are collected through an objective with $\text{NA}=0.6$. Right inset: relevant Yb transitions with dashed gray lines indicating transitions to dark states.
    Left inset: histogram of filling fraction for a 30-site tweezer array over 1000 realizations. (b) Mid-circuit readout scheme, single-shot image (top) and average (bottom) over 1000 realizations of the tweezer array with \SI{10.8}{\us} illumination time. The gray squares indicate the typical $3\times3$\,px ROIs for photon integration. (c) ROI photon counts histogram from $5000$ experimental images with \SI{10.8}{\us} illumination time. (d) Mean photon counts per atom versus illumination time. 
    Points are the array-average value and error bars (not visible) are the standard deviation. Dashed line: linear fit to the first 6 data points.} 
    \label{Fig.1}
\end{figure}

In our experiments, we prepare one-dimensional atomic arrays using optical tweezers at a wavelength of \SI{532}{\nm} and with a waist of \SI{580}{\nm}. We focus on ${}^{171}\text{Yb}$, an atom of central importance in quantum metrology, simulation and computing
~\cite{Gorshkov_2010,Ludlow_2015,Jenkins_2022,Ma_2022,Huie_2023,Ma_2023,Lis_2023,Norcia_2023}.
Single atoms are loaded almost deterministically from a narrow-line five-beam MOT~\cite{AbdelKarim_2025}, driving LACs with the MOT beams. 
We achieve $86.6(2)\%$ single-atom average filling for a 30-tweezers array [Fig.~\ref{Fig.1}(a)] with a trap depth of about \SI{570}{\uK}~\cite{SM}.
Imaging is performed by illuminating the atoms with alternated pulses from two counter-propagating beams resonant with the  ${}^1\text{S}_0\rightarrow{}^1\text{P}_1(F^\prime=3/2)$ transition at \SI{399}{\nm}. 
We employ trains of \SI{400}{\ns}-long pulses, short enough to mitigate the momentum transfer from a single beam~\cite{Bergschneider_2018,Su_2025}. 
The broad linewidth of the transition ($\Gamma\simeq2\pi\times29.1\,\text{MHz}$) and the high intensity ($s=I/I_\text{sat}\simeq 40$) of the imaging beams enable high-fidelity single-atom detection with microsecond-scale illumination time. 
Photons are collected by a $0.6\,\text{NA}$ microscope objective and are focused on a qCMOS camera, concentrating an atom's signal over few pixels.
We sum the camera photon counts into $3\times3$ pixels regions of interest (ROIs), obtaining well-separated histograms for empty and filled traps [Fig.~\ref{Fig.1}(c)].
The mean number of collected photons per atom increases linearly with the illumination time up to tens of microseconds [Fig.~\ref{Fig.1}(d)], in contrast with the count saturation previously observed with atoms escaping the ROI~\cite{Scholl_2023}.
Despite reduced momentum diffusion compared to continuous illumination~\cite{Su_2025}, pulsed illumination still induces recoil heating, requiring deep traps for confinement.
We measure a heating rate of \SI{19.4(4)}{\uK/\us} by release-and-recapture, consistent with expectations~\cite{SM}.
We note that traps operating at 532\,nm provide favorable confinement of the ${}^1\text{P}_1$ excited state, as they realize a near-magic condition for the ${}^1\text{S}_0\rightarrow{}^1\text{P}_1|F^\prime = 3/2, m_F^\prime = \pm3/2\rangle$ transitions and provide strong confinement of the $^1\text{P}_1|F^\prime = 3/2, m_F^\prime = \pm1/2\rangle$ states as well~\cite{SM}.

\begin{figure*}[t]
    \centering
    \vspace{0pt}
    \includegraphics[width=1\textwidth]{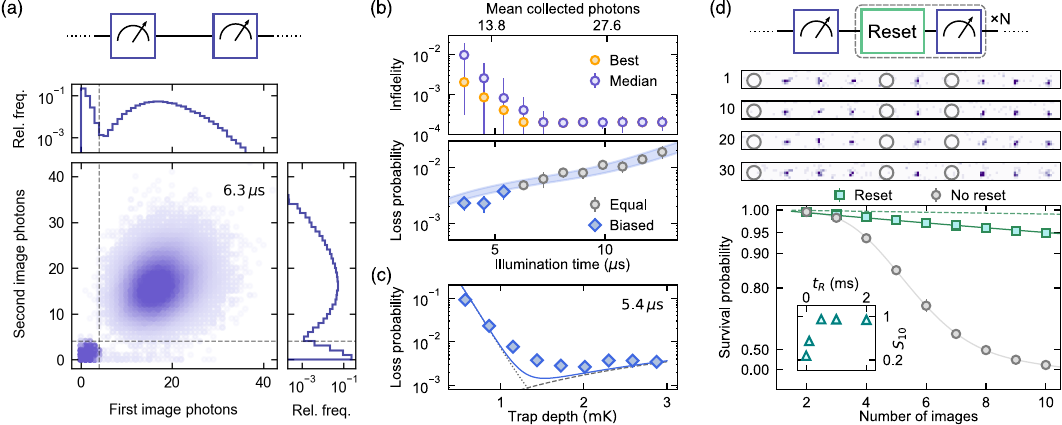}
    \caption{High-fidelity low-loss single-atom detection. (a) Top: Sequence used to measure imaging losses and detection fidelity. Bottom: correlation of photons collected in two consecutive images with illumination time $t_\mathrm{ill} = 6.3\,\text{\textmu s}$ and trap depth $U_0 \simeq2.27\,\text{mK}$. (b) Infidelity (top) and loss probability (bottom) versus illumination time and mean-photon counts. Purple and orange circles show the array median and best-tweezer infidelity. Gray circles and blue diamonds indicate the array-averaged loss probability from the equal-images and the fidelity-bias method, respectively. Shaded region: fit to a loss model with $2\sigma$ confidence interval. (c) Loss probability versus trap depth for $t_\mathrm{ill} = 5.4\,\text{\textmu s}$.
    The gray dotted and dashed lines show expected losses from heating and off-resonant scattering, respectively; the solid blue line combines both effects.
    (d) Top: Repeated imaging sequence with interleaved reset pulse. Center: single-shot images after 1, 10, 20 and 30 repetitions. Empty sites in gray. Bottom: survival probability versus number of images ($t_\mathrm{ill} = 6.3\,\text{\textmu s}$) without (gray) and with (green) reset pulse.
    The green solid (dashed) line is a fit of the measured survival after repeated detection-reset (reset alone) pulses, assuming uniform survival probability across repetitions. Gray line is a guide to the eye. Inset: survival probability after 10 images versus reset pulse duration for ${}^{174}\text{Yb}$. Data points in (b-c), (d) and inset are averaged over $5000$, $3000$ and $1000$ shots respectively. Error bars are the standard deviation across a 10-tweezers array.}
    \label{Fig.2}
\end{figure*}

We characterize the detection fidelity and the associated losses by acquiring two consecutive images of equal duration and analyzing their photon-count correlation.
Following a model-free approach~\cite{Holman_2024, Norcia_2018}, we extract the single-atom detection fidelity and survival probability, setting the photon-count threshold to maximize the fidelity. 
Fig.~\ref{Fig.2}(a) shows results for an illumination time $t_\text{ill} = 6.3\,\text{\textmu s}$, for which we achieve a fidelity of $99.96(4)\%$ and survival probability of $99.52(7)\%$ at a trap depth of \SI{2.27}{\mK}. 
As the average infidelity is dominated by a few underperforming tweezers, we report the median of the array as a representative figure of merit.
In contrast, the atom loss probability is uniform across the array, and we thus report the array-averaged value.
The shortest illumination employed ($t_\mathrm{ill} = 3.6\,\mu\mathrm{s}$) yields $99.0(9)\%$ fidelity and $99.77(6)\%$ survival probability, whereas longer illuminations improve the fidelity at the cost of a reduced survival [Fig.~\ref{Fig.2}(b)]. 
For $t_\mathrm{ill}\geq\,7.2\,\text{\textmu s}$ the fidelity saturates at $99.98(1)\%$, largely explained by decays from the ${}^1\text{P}_1$ to the triplet D-states~\cite{Loftus_2000}. 
For short illumination, the infidelity becomes comparable to or larger than the loss probability, making it increasingly difficult to reliably estimate the losses themselves. 
To overcome this issue, we adopt an asymmetric imaging sequence: we extend the illumination time of the second image to \SI{12.6}{\us} and we bias the threshold of the first image to higher values to suppress mislabeling of empty tweezers as occupied.
We then evaluate the loss probability $P_\text{loss}$ as the probability of detecting an atom in the first image but not in the second. 
We compare our results with a model accounting for different sources of losses, including imaging photon-recoil and trap-induced heating, decay to dark states and vacuum background collisions~\cite{SM}.
By fitting this model to the measured $P_\text{loss}$, we extract an imaging-induced heating rate of \SI{18.5(4)}{\micro\kelvin/\us}, consistent with an independent measurement (see above).
We investigate the dependence on trap depth at fixed $t_\mathrm{ill} = 5.4\,\text{\textmu s}$ and achieve $P_\text{loss}<1\%$ and trap-independent $99.9(1)\%$ fidelity, for all trap depths greater than $\sim$\,\SI{1}{\mK} [Fig.~\ref{Fig.2}(c)]. 
The loss probability saturates around $P_\text{loss}\simeq 0.3\%$, deviating from expectations of recoil heating and known decay to long-lived triplet P-states via D-states. 
We attribute excess losses to the off-resonant scattering of \SI{532}{\nm} trap photons, which induce excitations from the ${}^1$P$_1$ to higher-lying states as well as two-photon photoionization events. 
By treating the photoionization rate as the only free parameter, we obtain good agreement with the experimental data~\cite{SM}.
These findings suggest that trap photon scattering and photoionization are the dominant loss mechanisms for deeper traps. Operating at a longer far-off resonant trapping wavelength could further reduce these losses.

Although this method can be employed for repeated detections, motional excitations keep adding up leading to higher losses. 
%
%.
To mitigate this, we apply a cooling pulse driving the ${}^1\text{S}_0|F,m_F\rangle\rightarrow{}^3\text{P}_1|F^\prime = F+1,m_F^\prime = \pm m_F\rangle$ transitions to reset the atomic motional state after each image [Fig.~\ref{Fig.2}(d)].
Despite non-magic trapping, this reset pulse restores the initial temperature $\simeq$ \SI{20}{\micro\kelvin}, allowing to acquire over 30 repeated images with constant survival probability between successive images. 
Assuming a uniform loss probability across repeated detections, for $t_\mathrm{ill}=6.3\,\text{\textmu s}$ we observe a per-image survival probability of $99.43(1)\%$, only slightly lower than that of a single image.
This small reduction is likely caused by losses associated with each reset pulse, estimated as $0.096(8)\%$ from independent measurements~\cite{SM}. 
For the shortest illumination times, we verify that losses remain constant and compatible with those associated to a single detection for up to $30$ repetitions~\cite{SM}, while deviations from constant-loss behavior appear for $t_{\text{ill}}\gtrsim 8\,\text{\textmu s}$---likely due to suboptimal cooling rather than fundamental limits. 
In non-magic traps the Doppler cooling efficiency is compromised by differential light shifts that render the resonance condition energy-dependent.
We thus sweep the cooling-laser frequency towards resonance, enabling efficient cooling across different motional states.
In our setup, such sweep is limited to \SI{45}{\ms}, constrained by the laser frequency-locking bandwidth.
However, we show that orders-of-magnitude faster cooling is attained for ${}^{174}\text{Yb}$, for which trapping is magic on the ${}^1\text{S}_0\rightarrow{}^3\text{P}_1$ transition, eliminating the need of frequency sweeps.
In the inset of Fig.~\ref{Fig.2}(d) we display the survival probability after 10 images with $t_\mathrm{ill} = 6.3\,\text{\textmu s}$, demonstrating a performance comparable to ${}^{171}\text{Yb}$ with reset pulses as short as \SI{500}{\us}.
This enables a detection rate exceeding $ 2000\,\text{images}/\text{s}$ with a loss probability of $0.4(3)\%$ per image in $\sim$\,\SI{2}{mK} traps. 

\begin{figure}[b] 
    \centering
    \vspace*{-10pt}
    \includegraphics[width=\columnwidth]{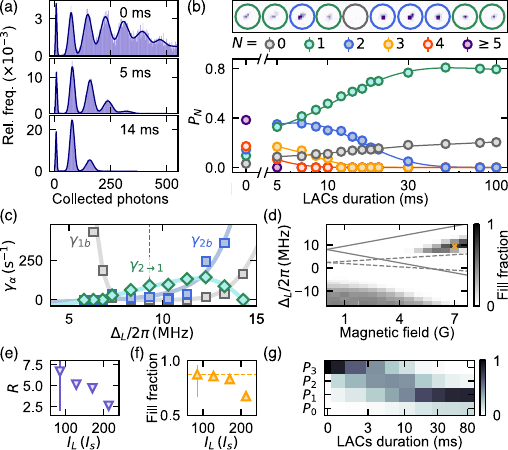}
    \caption{Number-resolved detection and LACs-driven near-deterministic loading dynamics. (a) Multi-atom photon-count histograms recorded after different LACs-pulse durations (0, 5 and 14\,ms), using $t_\mathrm{ill} = 20.7\,\text{\textmu s}$ and integrating over circular ROIs with $7$-px radius. Solid lines are multi-peak fits. (b) Top: single-shot image after a $7\,\text{ms}$ LACs pulse. ROI colors indicate the detected occupations. Bottom: filling probability versus blue-detuned LACs-pulse duration. Solid lines are fits with the rate-equation model~\cite{SM}. Error bars are not visible. 
    (c) Loss rates versus LACs-pulse detuning from the ${}^1\text{S}_0\rightarrow{}^3\text{P}_1(F' = 3/2)$ free-space resonance at fixed intensity $I_{L} = 170\,I_\text{sat}$. Solid lines are visual guides. Vertical dashed line marks the optimal detuning employed in the other panels. (d)~Filling fraction versus magnetic field and detuning for a $80\,\text{ms}$ LACs-pulse intensity $I_L\simeq215\,I_\text{sat}$. Dashed (solid) lines show the splitting of $m_F' = \pm3/2 (\pm1/2)$ states. The yellow cross marks the optimal parameters used in all other panels ($\Delta_L/2\pi = 9.4\,\text{MHz}$ and $7\,\text{G}$ magnetic field). (e) Good-to-bad collisions ratio $R$ as a function of LACs-pulse intensity. (f) Optimal filling fraction versus LACs-pulse intensity obtained from the rate equation fits; dashed line marks the maximum filling $\simeq 0.87$. Each point is obtained for the LACs-pulse duration that maximizes single-atom occupancy. (g) LACs dynamics with repeated multi-atom imaging. Data are obtained by post-selecting for $N=3$ initial occupation and correcting for losses induced by the first detection~\cite{SM}.    
    All data except in (d) are obtained from $\sim$1000 shots; data in (d) from 100~shots.}
    \vspace*{10pt}
    \label{Fig.3}
\end{figure}

Our fast imaging enables number-resolved detection in traps containing more than one atom~\cite{Grun_2024,Su_2025}. To prepare multiply-occupied traps, we shorten the LACs duration.
By extending the illumination time to \SI{20.7}{\us} and enlarging the camera ROIs to maximize photon counts, we resolve up to $7$ distinct peaks in the photon count histograms [Fig.~\ref{Fig.3}(a)]. 
We extract atom numbers by fitting histograms with a weighted sum of Gaussians, one for each occupation number, along with a Poissonian peak accounting for background counts, and we apply thresholding for classification~\cite{SM}.

We employ our atom number-resolved imaging to investigate the dynamics of ${}^\text{171}$Yb collisions in repulsive light-induced molecular potentials, leading to near-deterministic single-atom loading in tweezer traps \cite{Grunzweig_2010,Lester_2015,Brown_2019,Jenkins_2022}.
LACs rates have been previously characterized for rubidium~\cite{Fuhrmanek_2012,Weyland_2021,Sompet_2013,Pampel_2025} and erbium atoms~\cite{Grun_2024}, albeit with reduced atom-counting capability.
By measuring the evolution of the $N$-atom occupation probabilities $P_N$ for different LACs parameters, we extract loss rates using a rate-equation model involving up to four atoms~\cite{SM}. In particular, we determine the one-body ($\gamma_{1b}$), two-body ($\gamma_{2b}$) and two-to-one ($\gamma_{2\rightarrow1}$) loss rates~\cite{Pampel_2025}. 
The signature of near-deterministic loading is the dominance of $\gamma_{2\rightarrow1}$, which transfers population from $P_{N\geq 2}$ to $P_1$ without significantly increasing $P_0$ [Fig.~\ref{Fig.3}(b)]. 
We map the loss rates as a function of LACs-pulse detuning and intensity at a fixed magnetic field of \SI{7}{G}, shedding light on the dynamics of the near-deterministic loading region between the $m_F'=+1/2$ and $m_F'=+3/2$ resonances [Fig.~\ref{Fig.3}(d)], where $\gamma_{2\rightarrow1} \gg \gamma_{1b},\gamma_{2b}$.
At the optimal LACs-pulse detuning [Fig.~\ref{Fig.3}(d), yellow cross], the relative strength of $\gamma_{2\rightarrow1}$, parametrized by
$R=\gamma_{2\rightarrow{1}}/(\gamma_{1b}+\gamma_{2b})$,
and the maximum filling increase with decreasing LACs-pulse intensity [Fig.~\ref{Fig.3}(e,f)]. We conclude that a larger $R$ yields higher filling probabilities, despite a smaller $\gamma_{2\rightarrow1}$.
This is consistent with the picture of enhanced-loading resulting from a large number of low-energy collisions with small one-atom loss probability, but even smaller two-atom losses~\cite{Jenkins_2022}. 
In an alternative approach, we track LACs dynamics by acquiring two successive images with an interleaved LACs pulse and post-selecting on the number of atoms detected in the first image [Fig.~\ref{Fig.3}(g)]. 
Despite significant losses induced by the first image, we obtain consistent decay rates, verifying the absence of measurable three-body losses~\cite{SM}. 

\begin{figure}[t]  
    \centering
    \includegraphics[width=\columnwidth]{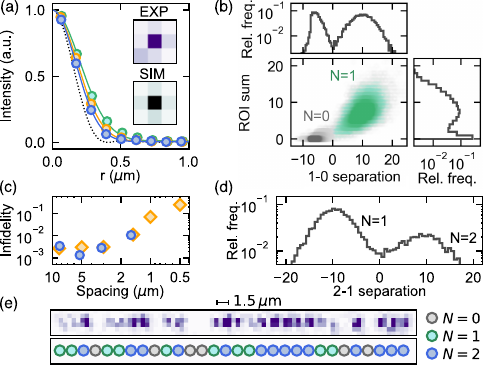}
    \caption{Fast microscopy of tightly-spaced arrays. (a) Size of the atomic signal from Monte Carlo simulations for different imaging conditions: $t_\mathrm{ill} = 5\,\text{\textmu s}$, $U_0=1.1\,\text{mK}$ (blue); $t_\mathrm{ill} = 20\,\text{\textmu s}$, $U_0=2.3\,\text{mK}$ (yellow); $t_\mathrm{ill} = 20\,\text{\textmu s}$, $U_0=1.1\,\text{mK}$ (green). Colored lines are Gaussian fits; black dotted line is the analytical PSF. Insets: experimental (top) and simulated (bottom) average signal from a single atom. (b) Comparison between photon counts in $1\times3$ ROIs and logL separation between filled and empty configurations~\cite{SM} for \SI{1.5}{\um}-spacing arrays with $t_{\mathrm{ill}}=5.4\,\text{\textmu s}$. %
    (c) Infidelity versus spacing for $t_{\mathrm{ill}}=5.4\,\text{\textmu s}$. Blue circles and yellow diamonds indicate experimental and simulated data respectively, obtained from $>10^4$ data points. Error bars not visible.
    (d) logL separation of the 2-1 atoms hypotheses for occupied sites in a multiply-filled array with \SI{1.5}{\um} spacing and $t_{\mathrm{ill}}=20.7\,\text{\textmu s}$. 
    (e) Single-shot MLE reconstruction with 0, 1 or 2 atoms per site in a 30-site array for $t_{\mathrm{ill}}=20.7\,\text{\textmu s}$.}
    \label{Fig.4}
\end{figure}

Unlike previous demonstrations of destructive fast imaging where the fluorescence signal spreads due to momentum-space diffusion~\cite{Ma_2023,Scholl_2023,Bergschneider_2018,Grun_2024,Su_2025}, the low-loss nature of our detection scheme ensures that atoms remain confined during imaging.
This substantially reduces image blurring and enhances spatial resolution.
In our setup, the fluorescence signal from individual atoms approaches the optical diffraction limit, a key requirement for extending this detection protocol to tightly-spaced arrays such as those employed in quantum gas microscopes.
We further confirm the high resolution and robustness of our approach through Monte Carlo simulations of the full imaging process, accounting for stochastic photon scattering and the associated atomic motion [Fig.~\ref{Fig.4}(a)]~\cite{SM}.

Leveraging the high spatial resolution of our detection technique, we explore its applicability to tightly-spaced arrays, making a first step towards its integration to lattice gas microscopes.
We prepare and image a 30-site array with a trap depth of \SI{1.1}{\milli \kelvin} and spacing of $\simeq$\,\SI{1.5}{\um}, where defining non-overlapping ROIs around each tweezer becomes infeasible. 
Under these conditions, conventional ROI-count thresholding fails and more advanced reconstruction methods---such as deconvolution algorithms~\cite{Omran_2015,LaRooij_2023,Buob_2024} and neural networks~\cite{Impertro_2023,Picard_2019}---are required. 
Here, we adopt a reconstruction algorithm based on maximum likelihood estimation (MLE).
For each experimental image, we compute the likelihood of all possible occupation-number configurations across the array by comparing the measured photon distribution to a model prediction, and select the most likely one~\cite{SM}.
As a figure of merit for single-atom detection, we compute the logarithm of the difference between the likelihood (logL separation) of the most likely configuration and that of alternative configurations in which the occupation of a single site is flipped.
In addition to photon-count statistics, the MLE approach also takes spatial information into account, enabling us to achieve high-fidelity detection in regimes where thresholding is not applicable [Fig.~\ref{Fig.4}(b)].
We assess the performance of this method by recording two equal images and evaluating the reconstruction fidelity as a function of tweezer spacing, applying the algorithm to simulated data for spacing beyond our experimental range [Fig.~\ref{Fig.4}(c)]. 
The reconstruction fidelity exceeds $99\%$, for spacing \SI{>1}{\um}. 
For sub-\SI{}{\um} spacing, fidelity degrades due to the limited spatial information caused by the low magnification in our current imaging system.
We further apply MLE to tightly-spaced arrays with doubly-occupied sites [Fig.~\ref{Fig.4}(e)]. 
In this case, as a figure of merit we use the likelihood separation between single and double occupancy, after having distinguished empty from occupied sites [Fig.~\ref{Fig.4}(d)]~\cite{SM}.
This approach could be extended to higher occupations of several atoms per site [see Fig.~\ref{Fig.3}(a)], at the cost of increased computational resources and refined modeling of the experimental spatio-temporal properties of the multi-atom signal. %---which can be directly calibrated experimentally. 

In conclusion, we have established a fast, minimally destructive imaging technique for single-atom detection in optical tweezers, achieving high fidelities and survival. 
We rapidly restore the atom motional state after detection, demonstrating a key capability for re-using atoms following mid-circuit qubit readout~\cite{Lis_2023,Norcia_2023,Bluvstein_2025,Zhang_2025b} or clock-state interrogation~\cite{Norcia_2019,Finkelstein_2024}.
Importantly, we have realized in-trap number-resolved  microscopy in dense arrays, without necessitating dynamical adjustment of atom-spacing~\cite{Koepsell_2020,Hartke_2020,Su_2023,Prichard_2024,Lebrat_2024,Boll_2016}---a novel functionality that can be incorporated in ytterbium lattice-gas microscopes in combination with advanced reconstruction algorithms. 
Such direct on-site atom counting, opens exciting prospects for quantum simulations of itinerant bosonic and fermionic lattice models, such as SU($N$)-symmetric~\cite{Taie_2012, Hofrichter_2016,Goban_2018, Tusi_2022, Pasqualetti_2024} and two-orbital Hubbard-like Hamiltonians~\cite{Gorshkov_2010,Riegger_2018,Ono_2021,Darkwah_2022}. 
The necessary mK-scale confinement can be obtained across $\sim1000$ sites in two-dimensional optical lattices, e.g.~using Watt-level laser power and leveraging bow-tie geometries or cavity-enhancement~\cite{Park_2022,Norcia_2024}.

Further, we have shed light on the dynamics of light-assisted collisions in gray molasses via repulsive molecular potentials, a fundamental mechanism underlying the near-deterministic loading of optical tweezer arrays. Our experiments will provide a benchmark for future theoretical efforts aimed at finding optimal regimes for loading efficiency and speed~\cite{Pampel_2025,Grun_2025}. 
We have also demonstrated first steps towards repeated multi-atom imaging, which could become a powerful tool for tracking population dynamics in a broad class of experiments. These results position fast in-trap imaging as a compelling approach for atom detection in quantum simulation, metrology and computing platforms.

\vspace*{20pt}

We thank M.~Aidelsburger, F.~Cesa, N.~Darkwah Oppong, F.~Ferlaino, S.~Jochim, P.~Lunt, G.~Pagano and J.~Thompson for insightful discussions. We also thank L.~Tanzi for lending us the qCMOS camera, and the Trieste Quantum initiative for the support. This work has received financial support from the European Research Council (ERC) under the European Union’s Horizon 2020 research and innovation programme (project OrbiDynaMIQs, GA No.~949438), and from the Italian MUR under the FARE 2020 programme (project FastOrbit, Prot.~R20WNHFNKF). This work has also received funding from the European Union under the Horizon Europe program HORIZON-CL4-2022-QUANTUM-02-SGA (project PASQuanS2.1, GA no.~101113690), and by the Next Generation EU (Missione 4, Componente 1) under the MUR PRIN 2022 programme (project CoQuS, Prot.~2022ATM8FY) and the PNRR MUR project PE0000023-NQSTI. 
R.K. acknowledge funding from the European Research Council (ERC) (Grant Agreement No. 101019739).

\end{document}

\clearpage
\setcounter{page}{1}

\onecolumngrid
\begin{center}
\textbf{Supplemental Material\\[8mm] 
\large Microsecond-scale high-survival and number-resolved detection of ytterbium atom arrays}\\[4mm]
A.~Muzi Falconi,$^{1}$ R.~Panza,$^{1,2}$ S.~Sbernardori,$^{1,2}$ R.~Forti,$^{1,3}$ R.~Klemt,$^{4}$ O.~Abdel~Karim,$^{2}$ M.~Marinelli,$^{1,5}$ F.~Scazza,$^{1,2,\ast}$\\[2mm]
\emph{\small $^1$Department of Physics, University of Trieste, 34127 Trieste, Italy}\\
\emph{\small $^2$Istituto Nazionale di Ottica del Consiglio Nazionale delle Ricerche (CNR-INO), 34149 Trieste, Italy}\\
\emph{\small $^3$Elettra Sincrotrone Trieste S.C.p.A., 34149 Trieste, Italy}\\
\emph{\small $^4$\,5.~Physikalisches Institut and Center for Integrated Quantum Science and Technology, Universit\"at Stuttgart, 70569 Stuttgart, Germany}\\
\emph{\small $^5$Department of Physics, University of Trieste, 34127 Trieste, Italy}\\
\emph{\small $^6$Istituto Officina dei Materiali del Consiglio Nazionale delle Ricerche (CNR-IOM), 34149 Trieste, Italy}\\
\end{center}

\vspace*{-10pt}

\begin{center}
\emph{\small ${}^\ast$ E-mail address: francesco.scazza@units.it}
\end{center}

\bigskip

\onecolumngrid

\section{Experimental details}

\begin{figure}[t]    \centering\includegraphics[width=0.98\columnwidth]{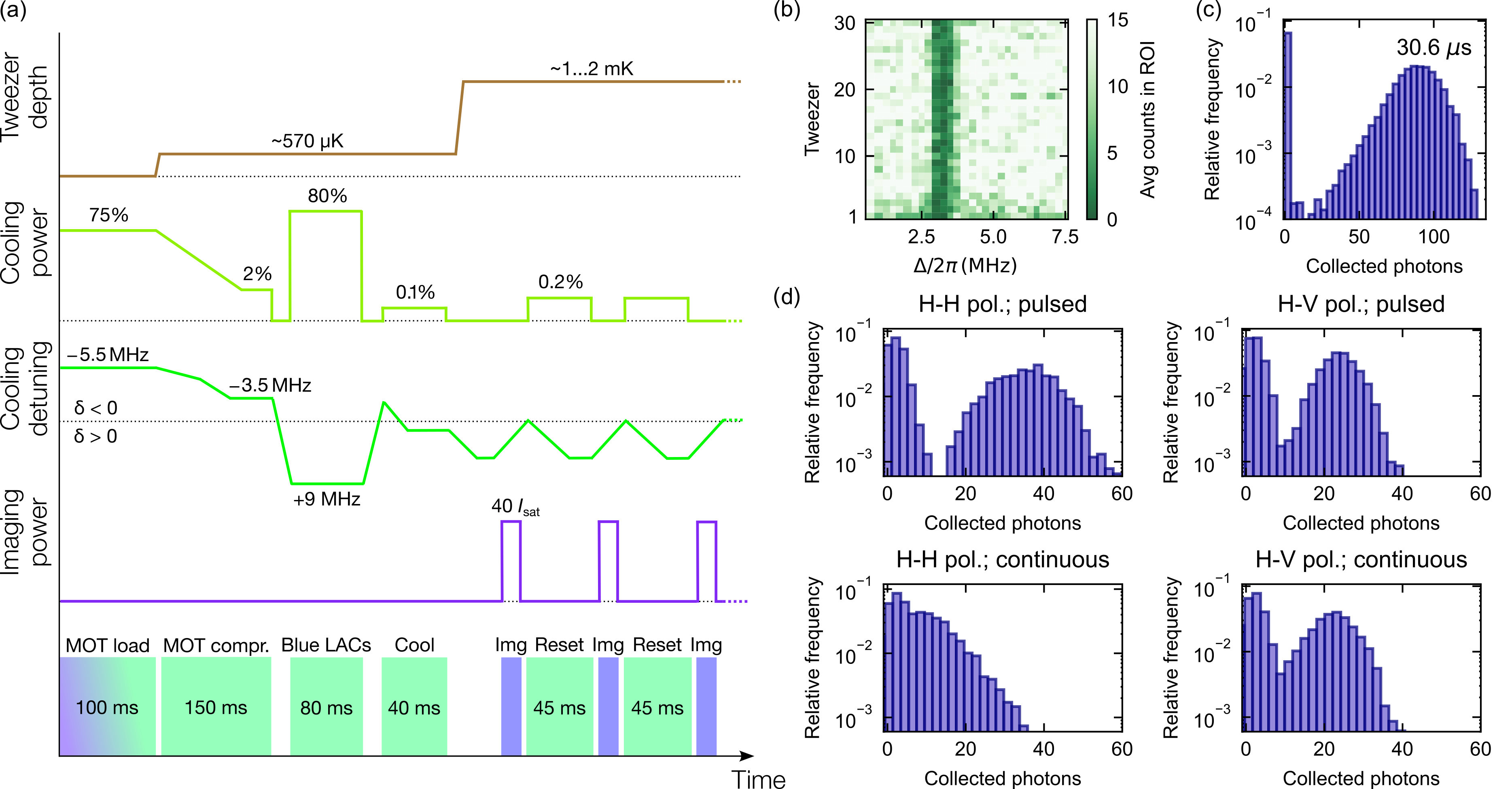}
    \caption{Additional experimental details. 
    (a) Scheme of the experimental sequence displaying the main steps for high-survival repeated imaging of ${}^{171}\text{Yb}$ arrays. $\delta$ denotes the detuning from the free-space ${}^1\text{S}_0|F=1/2\rangle\rightarrow{}^3\text{P}_1|F^\prime=3/2\rangle$ transition. Power and detuning levels are not to scale.
    (b) Site-resolved blow-out spectroscopy of the ${}^1\text{S}_0|F=1/2, m_{F} = \pm1/2\rangle\rightarrow{}^3\text{P}_1|F^\prime=3/2, m_{F}^\prime = \pm1/2\rangle$ transition in a 30-site tweezer array after trap intensity homogenization. Here, a trap depth of $U_0 \simeq 710\,\text{\textmu K}$ is used, and each data point is obtained from 25 shots. (c) ROI photon-count histogram from $5000$ experimental images with $t_\text{ill} = \SI{30.6}{\us}$. (d) Photon-count histograms for different polarization combinations of the two imaging beams (H: horizontal, V: vertical), and for pulsed or continuous illumination. Each histogram is obtained from 400 images with $t_\text{ill} = 20.7\,\text{\textmu s}$. The imaging trap depth for the histograms in panels (c) and (d) is $U_0 \simeq 2.3\,\text{mK}$.}  
    \label{SM Fig.1}
\end{figure}

A detailed description of our experimental setup, including the five-beam MOT loading scheme and the tweezer-array intensity homogenization procedure can be found in~\cite{AbdelKarim_2025}. 
We show a diagram of the experimental sequence used for producing and imaging single-atom arrays in Fig.~\ref{SM Fig.1}(a). 
In brief, for both ${}^{171}\text{Yb}$ and ${}^{174}\text{Yb}$, our experiments start with slowing and cooling an ytterbium atomic beam in a combined 2D MOT and Zeeman slower stage operating on the ${}^1\text{S}_0\rightarrow{}^1\text{P}_1$ transition. 
The slowed and collimated atom flux is then directed towards an octagonal glass cell, where atoms are captured and cooled in a 3D MOT operating with only five beams on the ${}^1\text{S}_0\rightarrow{}^3\text{P}_1$ transition. 
Additional crossed slowing beams operating on the ${}^1\text{S}_0\rightarrow{}^1\text{P}_1$ transition help reducing the speed of atoms incoming into the 3D MOT to below 10\,m/s, considerably increasing the loading rate.  
The MOT loading is followed by a MOT compression phase lasting approximately 150\,ms, where the magnetic B-field gradient is increased, and the cooling laser detuning and intensity are swept towards the resonance and down to the saturation intensity, respectively.
This allows to reach temperatures on the order of \SI{20}{\micro\kelvin}, measured by time-of-flight absorption imaging.
Optical tweezers are turned on at the beginning of the compression phase with a trap depth of about \SI{570}{\micro\kelvin}.
After the compression phase, the quadrupole B-field is turned off, and a light-assisted collisions (LACs) pulse is applied inducing single-atom occupations (near-deterministic for ${}^{171}\text{Yb}$).

The procedure to produce arrays of single ${}^{174}\text{Yb}$ atoms using red-detuned LACs has been described in~\cite{AbdelKarim_2025}. 
Here, we focus on ${}^{171}\text{Yb}$. 
We drive blue-detuned LACs employing the horizontal MOT beams with a detuning $\Delta_{L}/2\pi \simeq 9.4\,\text{MHz}$ from the free-space ${}^1\text{S}_0\rightarrow{}^3\text{P}_1 (F^\prime = 3/2)$ resonance with a magnetic field $B \simeq 7\,\text{G}$ aligned along the tweezers' polarization~\cite{Jenkins_2022}. 
LACs are performed at a trap depth of $\simeq570\,\unit{\micro\kelvin}$ to generate a differential (tensor) light shift of about 6\,MHz between the ${}^3\text{P}_1 \,|F = 3/2, |m_{F}|=3/2\rangle$ and $|F = 3/2, |m_{F}|=1/2\rangle$ states, which is the value that optimizes the single-atom fill factor.
The LACs pulse typically lasts \SI{80}{\ms} with an intensity of $\simeq215\,I_\text{sat}$, where $I_\text{sat} \simeq0.139\,\text{mW/cm}^2$ is the saturation intensity of the ${}^1\text{S}_0\rightarrow{}^3\text{P}_1$ transition.
After the LACs pulse, the magnetic field is reduced to \SI{0.2}{G} and single-atoms are Doppler cooled to \SI{20(3)}{\micro\kelvin} using a $40\,\mathrm{ms}$ red-detuned pulse on the ${}^1\text{S}_0|F=1/2, m_F = \pm1/2\rangle\rightarrow{}^3\text{P}_1|F^\prime = 3/2, m_{F}^\prime=\pm1/2\rangle$ transition.
The temperature in tweezers is measured via release-and-recapture~\cite{Tuchendler_2008}.
Also during this last cooling stage, the trap depth is kept at $\simeq570\,\unit{\micro\kelvin}$ where differential light shifts can be evened across the array.
After cooling, we keep the magnetic field constant while we adiabatically ramp up the trap depth before the fast imaging. The trap depth and magnetic field are then kept constant throughout the full imaging sequence as well as during the reset pulse.

Trap intensity homogenization is crucial for uniforming the efficiency of light-assisted collisions and cooling in non-magic traps. In Fig.~\ref{SM Fig.1}(b), we show an example of blow-out spectroscopy of the ${}^1\text{S}_0|F=1/2, m_{F} = \pm1/2\rangle\rightarrow{}^3\text{P}_1|F^\prime=3/2, m_{F}^\prime = \pm1/2\rangle$ transition across a 30-site tweezer array after a homogenization procedure~\cite{AbdelKarim_2025}. 
The spectroscopic signals from different tweezers (with a depth of $\simeq$ \SI{710}{\micro \kelvin} in this measurement) align with one another within the resonance width, and we measure a maximum deviation of $\simeq$ \SI{250}{kHz} between resonance frequencies across the array.

To image the atoms, we illuminate them with alternating pulses from two counter-propagating high-intensity ($I/I_\text{sat}\simeq40$) beams resonant with the ${}^1\text{S}_0\rightarrow{}^1\text{P}_1\,(F^\prime=3/2)$ transition. 
We typically detect single and multiple atoms with illumination times $t_\text{ill}\lesssim20\,\text{\textmu s}$. Longer illuminations are possible and yield similar high detection fidelities but more significant atom losses. The latter are responsible for the asymmetric tail visible in count histograms acquired with $t_\text{ill}\gtrsim 30\,\text{\textmu s}$ [Fig.~\ref{SM Fig.1}(c)].
The imaging beams are horizontally polarized (perpendicular to the tweezers polarization and the magnetic field), to maximize the photon collection efficiency by the microscope objective, due to the dipole emission pattern of atomic fluorescence~\cite{Su_2025}. 
Considering the objective aperture ($\text{NA} = 0.6$), the transmission through optical components and the qCMOS camera (Hamamatsu Orca Quest) quantum efficiency we estimate a total photon collection efficiency of $\simeq4.5\%$, a value consistent with photon counts obtained experimentally.
We compare histograms of collected photons for different imaging configurations to reveal the effect of the excitation beam polarizations and of the pulsed/continuous illumination on the imaging performance [Fig.~\ref{SM Fig.1}(d)]. We clearly observe a diminished performance for continuous illumination~\cite{Su_2025}, for both interfering (H-H) or non-interfering (H-V) imaging beams.
As expected, the pulsed H-V configuration is found to be worse than the pulsed H-H, since the emission pattern is optimally oriented for only one of the two excitation beams.
We therefore employ H-H pulsed beams for all our experiments.

The two imaging beams are obtained by splitting the output of a single laser source, offset-locked to our main \SI{399}{nm} master laser.
Each beam is focused and aligned through an AOM in order to minimize the pulse rise time down to around \SI{30}{ns} and coupled into a dedicated optical fiber.
Every element of the path, including fibers, cables as well as optical elements is chosen to be symmetric between the two paths.
The trains of alternated RF pulses driving the AOMs are generated by a FPGA (Red Pitaya STEMlab 124-14) with a custom firmware. 
The duration of each pulse is set to \SI{400}{ns} as a compromise between the limit imposed by the AOM rise time and the large momentum transfer from a single beam associated to longer pulses~\cite{Su_2025}. 
For the same reason, the first and last pulses of the train are set to \SI{200}{\nano\second} (half as all the others), and the length of illumination sequences (i.e., the values of $t_\text{ill}$) are chosen to ensure the same number of pulses per beam. 

As described in the main text, the operation of repeated low-loss imaging requires to add interleaved cooling pulses between subsequent images. To perform interleaved cooling of ${}^{171}\text{Yb}$, we chirp the cooling frequency towards resonance, addressing the reduced confinement of the excited state and enabling efficient cooling throughout different motional states.
In particular, we ramp the reset pulse detuning from $\Delta_{\text{R}}/2\pi \simeq -9.4\,\mathrm{MHz}$ to $\Delta_{\text{R}}/2\pi \simeq -2.4\,\mathrm{MHz}$ from the light-shifted \smash{${}^1\text{S}_0|F = 1/2, m_F = \pm1/2\rangle\rightarrow{}^3\text{P}_1 |F^\prime = 3/2, m_F^\prime = \pm1/2\rangle$} resonance and we employ a low intensity ($I_{\text{R}}\simeq 0.5\,I_{\text{sat}}$) [Fig.~\ref{SM Fig.1}(a)].
For cooling \smash{${}^{174}\text{Yb}$} in magic traps, we work at an intensity of $I_{\text{R}}\simeq 1.5\,I_{\text{sat}}$ with a constant detuning of $\Delta_{\text{R}}/2\pi \simeq -550\,\mathrm{kHz}$ from the (unshifted) \smash{${}^1\text{S}_0\rightarrow{}^3\text{P}_1 |J'=1,m_J=0\rangle$} resonance. 

We report in Table~\ref{Tab: isotopes comparison} a comparison between the parameters and results obtained for ${}^{171}$Yb and ${}^{174}$Yb for $t_\text{ill}=6.3\,\text{\textmu s}$ and trap depth $\simeq\,2\,$mK. 
We find very similar results for the two isotopes, the only noticeable difference being a slightly higher mean photon count for ${}^{174}\text{Yb}$.
\begin{table}[h!]
    \bigskip
	\centering\small
    \setlength{\tabcolsep}{8pt}
	\begin{tabular} {lc|c}
	\toprule
	& \multicolumn{1}{c|}{${}^{171}\text{Yb}$} & \multicolumn{1}{c}{${}^{174}\text{Yb}$}  \\ 
		\midrule
		\midrule
        Mean photon counts & 17(2) & 19(2) \\
        Detection fidelity & 99.96(4)\% & 99.98(1)\% \\
        Survival probability & 99.52(7)\% & 99.67(1)\% \\
        Reset pulse duration & 35\,ms (sweep) + 10\, ms & 0.5\,ms \\
        Reset pulse detuning & $-9.4\,\text{MHz}\rightarrow-2.4\,\text{MHz}$ & -0.55\,MHz \\
        Reset pulse intensity & 0.5\,$I_\text{sat}$ & $1.5\,I_\text{sat}$ \\
        Per-image survival (10 images) & 99.43(1)\% & 99.4(1)\% \\
        \bottomrule
	\end{tabular}
	\caption{Comparison between the imaging parameters and performance of ${}^{171}\text{Yb}$ and ${}^{174}\text{Yb}$ for 6.3\,\textmu s illumination time and trap depth $\simeq 2\,$mK.}
	\label{Tab: isotopes comparison}
\end{table}

\section{Extraction of imaging fidelity and loss probability}

\begin{figure}[t]    
\centering\includegraphics[width=0.5\columnwidth]{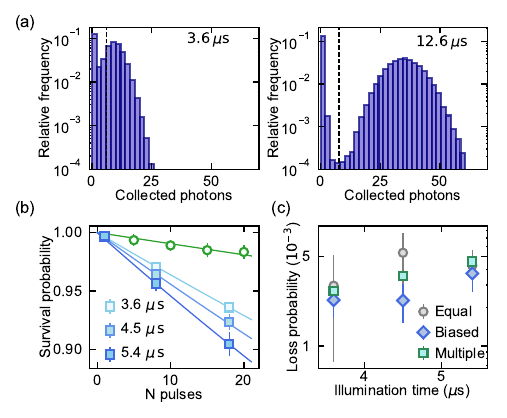}
    \caption{Imaging loss analysis. (a) Histograms of the ROI photon counts for the asymmetric imaging sequence. Left: histogram of the first image with short illumination time (here $t_\text{ill}=3.6\,\text{\textmu s}$). Right: histogram of the second image with longer illumination time ($t_\text{ill}=12.6\,\text{\textmu s}$). Dashed lines indicate the photon count thresholds for single-atom detection. Each histogram is obtained from 5000 experimental images. (b) Survival probability after a variable number of pulses. Green circles: effect of the reset pulses alone. Blue squares: interleaved detection and reset pulses; darker colors indicate longer illumination times. Solid lines are fits to the data assuming a constant loss probability per pulse. (c) Comparison between loss-probabilities extracted with different methods for the shortest illumination times. Green squares indicate the extracted single pulse loss from the fits in (b). All methods yield compatible losses. Each point is obtained from $>3000$ experimental images and error bars are standard deviation across the array. For all panels the imaging trap depth is \SI{2.27}{\milli \kelvin}.}  
    \label{SM Fig.2}
\end{figure}

We estimate the fidelity and losses associated to our imaging following the model-free approach presented in Refs.~\citenum{Holman_2024, Norcia_2018}. We acquire two equal images separated by about \SI{25}{ms}, and, by analyzing many realizations, we determine for each site of the array the probability of measuring an atom in both the images ($P_{11}$), a void in both the images ($P_{00}$), an atom and then a void ($P_{10}$) or a void and then an atom ($P_{01}$). The probability of each outcome is given by:
\begin{align*}
    & P_{11} = f(1-F_0)F_1(1-S)+(1-f)(1-F_0)^2+fF_1^2S \\[5pt] & P_{10} = fF_1S(1-F_1)+fF_1(1-S)F_0 + (1-f)(1-F_0)F_0 \\[5pt] 
    & P_{01} = f(1-F_0)(1-F_1)(1-S)+(1-f)(1-F_0)F_0 + f(1-F_1)F_1S \\[5pt] 
    & P_{00} = 1-(P_{11} + P_{10} + P_{01})
\end{align*}
where $f$ is the filling fraction, $S$ is the survival probability after one image and $F_1$ and $F_0$ are the fidelities of correctly identifying an atom or a void, respectively. The total single-atom detection fidelity is defined as $F = fF_1+(1-f)F_0$. For each trap, we solve the equations above for different photon-count threshold values and we choose the value that maximizes the total fidelity. 
As our detection scheme injects motional excitations into the system, the initial state of the atoms in the second image is indeed different compared to the first.
However, we do not observe any significant effect of the first detection on the second image photon-count histogram for most of the investigated illumination times, with only minor effects for the longest illumination times ($t_{\mathrm{ill}}\gtrsim10\,\text{\textmu s}$). 
We also verify that introducing an interleaved cooling pulse between the images does not lead to significant improvements in the measured fidelity and survival probability. 

For the shortest illumination times ($t_{\mathrm{ill}}\lesssim 5\,\text{\textmu s}$) we estimate the survival probability using an asymmetric imaging sequence, where the threshold of the first image is biased to larger values and the illumination time of the second image is increased to \SI{12.6}{\us} [Fig.~\ref{SM Fig.2}(a)].
We then compute the loss probability simply as the probability of detecting an atom in the first image and a void in the second. 
This approach yields results compatible with the equal-duration images method, while being more robust to infidelities [Fig.~\ref{SM Fig.2}(c), circles and diamonds]. 
The measured loss probabilities also match those obtained from fitting the per-image loss probability in multiple, repeated detection-reset sequences [Fig.~\ref{SM Fig.2}(b,c), squares], correcting for the $0.096(8)\%$ loss caused by each reset pulse.
To obtain the per-pulse losses we fit the data with $S_{N} = \left(1-P_\text{loss}\right)^N$, where $S_N$ is the survival probability after $N$ pulses and $P_\text{loss}$ is the per-pulse loss probability.

\section{Imaging loss mechanisms}

\subsection{Trapping of excited ${}^1\text{P}_1$ state}
\begin{figure}[t]    \centering\includegraphics[width=0.55\columnwidth]
{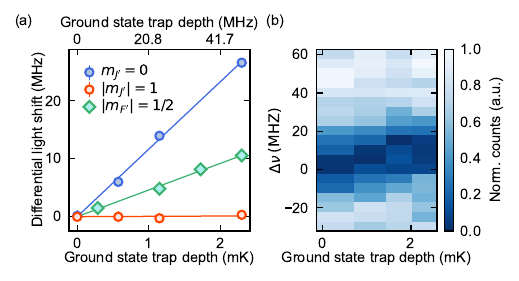}
    \caption{${}^1\text{P}_1$ excited state light shifts. (a) Differential light shift between the ground and excited state versus ground state trap depth for different sublevels of ${}^{174}\text{Yb}$ (circles) and ${}^{171}\text{Yb}$ (diamonds). Solid lines are linear fits to the data. (b) Blow-out spectroscopy of ${}^{171}\text{Yb}$ employing $\sigma$-polarized (same polarization as the imaging beams) probe beams to excite all allowed transitions.}  
    \label{SM Fig.3}
\end{figure}

We verify that the excited state of the imaging transition is trapped in \SI{532}{\nm} tweezers, ruling out the anti-trapping of ${}^1\text{P}_1$ as a possible loss source.
We measure the differential light shift between the ground and excited state of ${}^{174}\text{Yb}$ in $\pi$-polarized tweezers by performing blow-out spectroscopy of the ${}^1\text{S}_0\rightarrow{}^1\text{P}_1$ transition. 
Opposite to the situation for the ${}^3\text{P}_1$ state, we find that the ${}^1\text{P}_1|J^\prime=1, m_{J}^\prime = 0\rangle$ state is significantly less trapped than the ground state, with a measured differential light shift of \SI{11.6(2)}{MHz/\milli \kelvin} [Fig.~\ref{SM Fig.3}(a), blue circles]. 
Strikingly, our measurements reveal that the ${}^1\text{S}_0 |J=0,m_{J}=0\rangle \rightarrow{}^1\text{P}_1 |J^\prime=1,m_{J}^\prime=\pm1\rangle$ transitions feature instead a vanishing differential light shift for $\pi$-polarized trapping light at \SI{532}{\nm}, indicating a near-magic trapping condition [Fig.~\ref{SM Fig.3}(a), red circles].
For these stretched transitions, the differential polarizability is independent of the isotope, i.e., it is expected to always equal the one measured in ${}^{174}\text{Yb}$. 
Therefore, we expect the ${}^1\text{S}_0|F=1/2,m_{F}=\pm1/2\rangle\rightarrow{}^1\text{P}_1|F^\prime=3/2,m_{F}^\prime=\pm3/2\rangle$ in ${}^{171}\text{Yb}$ transitions, which are the strongest transitions addressed by our imaging beams, to be unshifted by our trapping light. 
By selectively addressing ${}^1\text{S}_0|F=1/2,m_{F}=\pm1/2\rangle\rightarrow{}^1\text{P}_1|F^\prime=1/2,m_{F}^\prime=\pm1/2\rangle$ transitions in ${}^{171}\text{Yb}$ with $\pi$-polarized blow-out beams, we measure the differential polarizability of the involved states, finding \SI{4.6(1)}{MHz/\milli \kelvin} [Fig.~\ref{SM Fig.3}(a), green diamonds]. 
On the other hand, we cannot selectively address the stretched transitions in our setup at low B-fields, as the tensor light shift is not sufficient to split levels with different values of $|m_{F}^\prime|$ due to the broad linewidth of the transition. We therefore cannot isolate the differential light shift of the $m_{F}^\prime = \pm3/2$ states alone.
Figure~\ref{SM Fig.3}(b) displays the result of blow-out spectroscopy employing a beam with the same polarization as our imaging light, i.e., $\sigma$-polarization, where all allowed transitions within the $\sim$ 30\,MHz linewidth are addressed. 
The weak dependence of the resonant frequency on ground state trap depth is mostly caused by excitations of the $m_{F} =\pm1/2\rightarrow m_{F}^\prime = \mp1/2$ transitions, and it is another strong indicator of the vanishing differential light shift of the $m_{F}^\prime = \pm3/2$ states. 
Our findings suggest that the small differential light shift between the ground and excited state is highly beneficial to suppress atom-loss during detection, as it provides strong confinement of the excited state as well as no additional heating from dipole force fluctuations~\cite{Martinez_2018}.

\subsection{Loss model}
To gain insight into the mechanisms leading to the measured imaging losses, we compare experimental data with a theoretical model taking into account different atom-loss sources. In particular, we estimate the loss probability as:

\begin{equation}
    \label{eq: all losses}
    P_\text{loss} = P_\text{h,v}(U_\text{eff},T_\text{ill},t_\text{ill}) + P_\text{h,v}(U_0,T_\text{wait},t _\text{wait}) 
    + P_\text{triplet}  + P_\text{trap,1} + P_\text{trap,2}
 \end{equation}   
where $P_\text{h,v}(U,T, t)$ is the heating- and vacuum-induced loss probability after a time $t$ for an atom with temperature $T$ (which itself increases over time $t$) in a trap of depth $U$; $U_\text{eff}$ is the average effective trap depth experienced by atoms spending part of the illumination time ($t_\text{ill}$) in the ground and part in the excited state; $U_0$ is the ground state trap depth; $t_\text{wait}$ is the wait time between consecutive images; $T_\text{ill}$ is the atomic temperature after imaging; $T_\text{wait}$ is the atomic temperature after the wait time between consecutive images; $P_\text{triplet}$ is the loss probability associated to decays to long-lived triplet states; and $P_\text{trap,1}$ and $P_\text{trap,2}$ account for losses caused by one and two-photon off-resonant scattering from the ${}^1\text{P}_1$ excited state, respectively, from the trapping laser.

We estimate heating- and vacuum-induced losses as~\cite{Tuchendler_2008}: 
\begin{equation} 
\label{eq: tuchendler} 
P_{\text{h,v}}(U,T, t) = 
1 - \left[ 1-\left(1+\eta + \frac{1}{2}\eta^2\right)e^{-\eta}\right]e^{-t/\tau_\text{v}}
\end{equation}
where $\tau_\text{v}\simeq180\,\text{s}$ is the vacuum lifetime and $\eta = U/k_BT$. 
The atomic temperature increases linearly with time due to both the imaging photon-recoil heating rate $\alpha$ and by the trap-induced heating rate $\beta\simeq60\,\text{\textmu K/s}$ (indipendently measured), caused by off-resonant scattering and technical noise. To estimate $\alpha$, we perform release-and-recapture~\cite{Tuchendler_2008} measurements after variable illumination time, finding $\alpha=19.4(4)\,\text{\textmu K/\textmu s}$, consistent with expectations from the photon recoil-induced increase of thermal energy~\cite{Grimm_2000} and with results from fitting Eq.~\eqref{eq: all losses} to the experimental data. Considering both the heating rates we have: $T_\text{ill} = T_0+\alpha\,t_\text{ill}+\beta\,t_\text{ill}$ and $T_\text{wait}=T_\text{ill}+\beta\,t_\text{wait}$, where $T_0\simeq20\,\text{\textmu K}$ is the initial atomic temperature. Losses caused by heating and vacuum background collisions are therefore accounted for by the first two terms of Eq.~\eqref{eq: all losses}. 

Atoms can be lost from the imaging cycle due to spontaneous ${}^1\text{P}_1\rightarrow^3\text{D}_{1,2}$ decays. 
From the $^3\text{D}_{1,2}$ states, atoms can either decay to the metastable ${}^3\text{P}_{0,2}$ states, resulting in an effective atom-loss, or decay first to the $^3\text{P}_1$ state and then back to the ground state with a typical timescale of $\tau_{{}^3\text{P}_1} \simeq 0.86\,\text{\textmu s}$, leading to a detection infidelity. 
Compared to other species possessing a similar electronic structure, such as alkaline-earth atoms or lanthanides, the \smash{${}^1\text{P}_1\rightarrow{}^3$D$_{1,2}$} decay rates of ytterbium are extremely low~\cite{Porsev_1999,Loftus_2000,Cho_2012}, leading to a very low probability $P_\text{triplet}\approx 10^{-7}\,N_\text{ph}$ of decaying into long-lived triplet states, where $N_\text{ph}$ is the number of scattered imaging photons. 

Notably, we find that the most relevant loss source, beside imaging-induced heating, is the off-resonant scattering of trap photons from the ${}^1\text{P}_1$ state.  
Employing excited level lifetimes from~\cite{Blagoev_1994} as inputs, we estimate the expected \SI{532}{nm} scattering rate for an atom in the ${}^1\text{P}_1$ state in a \SI{1}{\milli\kelvin}-deep trap as $a_\text{1-ph}\simeq 115\,\text{s}^{-1}\text{mK}^{-1}$, dominated by a relatively broad ($\Gamma\simeq2\pi\times3.2\,\text{MHz}$) transition to the 6s7d${}^1\text{D}_2$ state at \SI{518.4}{\nm}. As this state is anti-trapped by the tweezers light, we estimate the loss probability due to off-resonant trap photon scattering as $P_\text{trap, 1} = N_\text{ph}\,a_\text{1-ph}\,U_0\,\tau_{{}^1\text{P}_1}$, where $\tau_{{}^1\text{P}_1}\simeq5.46\,\text{ns}$ is the ${}^1\text{P}_1$ state lifetime.
Finally, we take into account the loss probability due to two-photon photoionization events from the ${}^1\text{P}_1$ state as $P_\text{trap,2} = N_\text{ph}\,a_\text{2-ph}\,U_0^2\,\tau_{{}^1\text{P}_1}$. We estimate the photoionization rate per mK from our experimental data, finding $a_\text{2-ph}\simeq180\,\text{s}^{-1}\text{mK}^{-2}$, on the same order of magnitude of extracted two-photon photoionization rates from triplet states in \SI{487}{nm} traps~\cite{Ma_2023}. 
We believe that the largest contribution to this rate is caused by the 6s7d${}^1\text{D}_2$ intermediate state.
Even though any trapping wavelength shorter than $\sim$\,\SI{800}{\nm} is sufficient to induce two-photon photoionization events, their probability could be strongly suppressed by employing trapping wavelengths near or above $\sim$\,\SI{760}{\nm}, given that the nearest transition (${}^1\text{P}_1\rightarrow 6s6d{}^1\text{D}_{2}$) has a wavelength of 667\,nm.

We do not find any dependence of the measured losses on the imaging light intensity for fixed number of scattered photons, and we thus exclude that two-photon excitation processes driven by imaging photons play a significant role.

\section{Multi-atom counting analysis}
We employ two different approaches to fit the photon-count histograms in the presence of multiple atoms per tweezer.
The first one is agnostic to the number of trapped atoms, and it is used to determine the thresholds for single-shot atom counting.
The second one is used to extract the occupation-number probabilities to study the LACs dynamics.

In the first case, we fit our photon-count histograms with a multi-peak fitting function defined in the following way: the dark (zero-atom) peak is fitted with a Poissonian distribution, while the bright peaks corresponding to discrete atom numbers with a sum of Gaussian functions. In addition, we employ an exponential tail to account for high-count events, where individual peaks are no longer well-resolved. 
The full fitting function reads:
\begin{equation*}
    f_{N}(n) = A_0\,\frac{x_0^n}{n!}e^{-x_0} \, + \, \sum_{i=1}^{N-1} A_i \,\exp\left(-\frac{(n-x_i)^2}{2\sigma_i^2}\right) \, + A_N \,(n-x_N)^2\, e^{-\gamma(n-x_N)}\,\theta(n-x_N)
\end{equation*}
where $N$ is the total number of considered peaks, $n$ are the photon counts, $x_i$, $\sigma_i$ and $A_i$ are the center, width and amplitudes of the Gaussian peaks, $\gamma$ is a decay constant and $\theta$ is the Heaviside function. 
The last term of the fitting function is designed to optimize the fit of the last peak before the exponential tail. 
Accurate fitting requires careful setting of the initial guess for the various parameters (e.g., peak positions and amplitudes), which we implement via peak-search on a smoothed version of the histogram.

From the fit parameters, we define thresholds to distinguish between $k$ and $k+1$ atoms: each threshold $n_k$ is obtained by minimizing the discrimination infidelity defined as the probability of having $n<n_k$ ($n>n_k$) counts for $k+1$ ($k$) atoms:
\begin{equation*}
    I_d(n_k) = \frac{1}{2}\int_{-\infty}^{n_k}\exp\left(-\frac{(n-x_{k+1})^2}{2\sigma_{k+1}^2}\right) + \frac{1}{2}\int^{+\infty}_{n_k} \exp\left(-\frac{(n-x_k)^2}{2\sigma_k^2}\right)
\end{equation*}
These thresholds allow us to count up to $N = 6$ atoms per site in each image [see Fig. 3 of the main text]. We extract the counting fidelity as done in~\cite{Grun_2024}, obtaining fidelities of \SI{99.6}{\percent}, \SI{97.6}{\percent}, \SI{94.2}{\percent}, \SI{90.3}{\percent}, \SI{86.1}{\percent}, \SI{81.3}{\percent} for counting $N = 1,2,3,4,5,6$ atoms, respectively.
The fidelity decreases for larger atom number due to the broadening of the Gaussian distributions for higher numbers of collected photons.

To study the LACs dynamics with a single multi-atom image we employ a second type of fitting.
Unlike the previous method, which uses a non-normalized fitting function, for this second approach we use a normalized model. 
Here, we restrict our analysis to $N\leq4$ atoms, for which  histogram peaks are well separated and fidelities exceed $90\,\%$ using $t_\text{ill} = 20.7\, \text{\textmu s}$.
We apply this method to histograms displaying a negligible probability of $N>4$ events.
The fitting function reads:
\begin{equation*}
    f_{N}(n) = \,P_{0,p}\,\mathcal{P}(x_{0,p};\,n)\, + \, P_{0,g}\,\mathcal{N}(x_{0,g},\,\sigma_0;\,n) + \,  \sum_{i=1}^{N-1} P_i\,\mathcal{N}(x_{i},\,\sigma_i;\,n)  + \,  \left(1-\sum_{i=0}^{N-1}P_i\right) \,\mathcal{N}(x_{N},\,\sigma_N;\,n)
\end{equation*}
where $\mathcal{P}(x;\,n)$ is the value at $n$ of a Poisson distribution centered in $x$, $\mathcal{N}(x,\,\sigma;\,n)$ is the value at $n$ of a normal distribution centered in $x$ with width $\sigma$, and $P_i$ is the probability of having a trap containing $N=i$ atoms. 
For the dark peak, the fit combines a Poisson and a normal distribution, and the probability of having an empty trap is then $P_0=P_{0,p}+P_{0,g}$.
This combination increases the fitting accuracy in larger ROIs featuring Gaussian noise that deforms the Poissonian peak. The probabilities obtained from this fitting routine have been used to extract the parameters of the rate equations describing the LACs dynamics (see Fig. 3 and Sec. V).

\section{Rate equation modeling of light-assisted collision dynamics}
\subsection{Model}
To fit the time evolution of site occupancy under the effect of the LACs pulse, we use a system of coupled rate equations.
The model includes three loss mechanisms induced by the LACs light: a one-body loss rate $\gamma_{1b}$ due to motional heating, a two-body loss rate $\gamma_{2b}$, and $\gamma_{2\rightarrow 1}$ corresponding to the loss a single atom in a pair~\cite{Grunzweig_2010,Brown_2019,Jenkins_2022,Pampel_2025}. 
The rate equations are written as:
\begin{equation*}
    \begin{cases}
        \dot{P_0}(t) = \gamma_{1b} \, P_1(t) + \gamma_{2b} \, P_2(t) \\
        \dot{P_1}(t) = -\, \gamma_{1b} \, P_1(t) + (\,\gamma_{2\rightarrow 1} + 2\, \gamma_{1b} ) \, P_2 (t)  +\,3\, \gamma_{2b} \, P_3(t) \\
        \dot{P_2}(t) = - (\,\gamma_{2b}+\gamma_{2\rightarrow 1} + 2\, \gamma_{1b}) \, P_2(t) +\,3\,(\gamma_{2\rightarrow 1} + \gamma_{1b} ) \, P_3 (t) + 6\,\gamma_{2b} \, P_4(t) \\
        \dot{P_3}(t) = - 3\,(\gamma_{2b}+\gamma_{2\rightarrow 1} + \gamma_{1b}) \, P_3(t)+\,(6\,\gamma_{2\rightarrow 1} + 4\,\gamma_{1b} ) \, P_4 (t)\\
        \dot{P_4}(t) = - (6\, \gamma_{2b}+6\, \gamma_{2\rightarrow 1} + 4\, \gamma_{1b}) \, P_4(t)\\
        P_4(t) = 1 - \left[P_0(t) + P_1(t) + P_2(t) + P_3(t)\right]
    \end{cases}
\end{equation*}
    where $P_N(t)$ is the probability of having $N$ atoms at time $t$; the last equation accounts for the normalization condition. 
    The integer prefactors account for the number of equally-likely ways to loose one or two atoms among the $N$ available. 
    We limit our model to 4 atoms per site, as the peak fitting becomes less reliable for higher occupations.
    To ensure the validity of this truncation, we limit our fitting of the probability evolutions to times for which the probability $P_{\geq5}(t)$ of having $N\geq5$ atoms is negligible.

In the experiment, we prepare tweezers containing multiple atoms at a depth of $\simeq570\, \unit{\micro\kelvin}$. In the case of repeated multi-atom detection, we then raise the trap to \SI{2.27}{\milli\kelvin} and apply an initial imaging pulse for \SI{10.8}{\micro\second} to enable post-selection of the number of atoms considered in the ensuing dynamics. 
A LACs pulse is then applied with a variable duration while holding atoms in $570\, \unit{\micro\kelvin}$-deep traps.
After this pulse, we image the array in-trap at a depth of $2.27 \,\unit{\milli\kelvin}$, with $t_\text{ill}=20.7\,\text{\textmu s}$ for the single image or $t_\text{ill}=10.8\,\text{\textmu s}$ in the case of repeated images.
For the single image case we fit the histograms corresponding to each LACs pulse duration and extract the atom-number probabilities as described in sec. IV.
To extract the loss rates, we first numerically integrate the coupled rate equations system using the \texttt{integrate} package of \texttt{scipy} and then fit the rate equations with the three loss rates $\gamma_{1b}$, $\gamma_{2b}$ and $\gamma_{2\rightarrow 1}$ as free parameters.
For this purpose, we minimize the sum of the square residuals between the model and the data, weighted using the experimental error bars, using the Python package \texttt{lmfit}.

To exclude the relevance of higher-order processes we tested the addition of a three-body loss rate, associated to three-body collisions, to the rate equations.
The resulting fit yielded a cost function value an order of magnitude higher than in the case without the three-body loss rate and we therefore concluded that three-body losses do not play a significant role in our dynamics.
Moreover, we find that the near-deterministic regime is characterized by a relatively small two-to-one rate, albeit much larger than the one- and two-body loss rates. This supports the interpretation that enhanced loading results from multiple collisions, each more likely to remove atoms one at a time rather than in pairs, as previously suggested in~\cite{Jenkins_2022}. 
On the other hand, for detunings close to the Zeeman-shifted transitions [Fig.~3(c)], the dominant loss source are one-body losses and fitted values of $\gamma_{2b}$, and $\gamma_{2\rightarrow 1}$ are compatible with zero.
In this regime, we repeat the fitting procedure with these two rates set to zero.
The cost function value remains approximately unchanged and the reduced number of fitting parameters allows for a more precise determination of the one-body loss rate. 

\subsection{Repeated multi-atoms imaging}
We also apply the rate equations model to the case of repeated images.
In this case, there are no assumptions on the maximum number of atoms as post-selection in the first image allows to know exactly the initial number of atoms to be considered in the rate equations.
On the other hand, the lack of an efficient cooling prior to the first multi-atom image leads to significant losses after the first detection. 

\begin{figure}[t]    \centering\includegraphics[width=0.65\columnwidth]{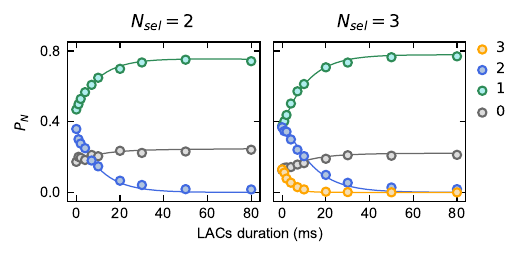}
    \caption{Filling probability dynamics in blue-detuned LACs for post-selected repeated images with $N_{sel}=2$ and $N_{sel}=3$. Solid lines are fits with the rate equations model and different colors correspond to filling probability for different atom numbers in the second image. Each point is obtained from more than 2200 experimental shots.}  
    \label{SM Fig.4}
\end{figure}

By post-selecting traps with an initial population  $N_\text{sel}=2$ and $N_\text{sel}=3$, and fitting their evolution, we are able to confirm both the loss rate values obtained with the single-image approach and the absence of a measurable three-body loss process [Fig.~\ref{SM Fig.4}]. 
The extracted loss rates at $I_L\sim150\,I_\text{sat}$ are consistent with the results obtained using single-image measurements: for $N_\text{sel}=2$, $\gamma_{1b} = 0.1(2)\,\unit{\per\second}$, $\gamma_{2b} = 18(2)\,\unit{\per\second}$ and $\gamma_{2\rightarrow1} = 76(3)\,\unit{\per\second}$, while for $N_\text{sel}=3$, $\gamma_{1b} = 0.05(17)\,\unit{\per\second}$, $\gamma_{2b} = 17(1)\,\unit{\per\second}$ and $\gamma_{2\rightarrow1} = 71(1)\,\unit{\per\second}$. 
In the case of $N_\text{sel}=3$, it is also possible to add a three-body loss process to the $\dot{P_3}$ term through a rate $\gamma_{3b}$. 
We find a very small value for such rate ($\sim10^{-7}\,\unit{\per\second}$), consistent with the absence of such loss mechanism.
By comparing data and the rate equations model for $N_\text{sel}=2$ and $N_\text{sel}=3$, the achieved final $N=1$ filling is slightly larger for the latter case, indicating that loading tweezers with $N\gg2$ (to keep the probability of starting with $N=2$ as low as possible) before triggering LACs is beneficial to achieve the largest fraction of one-atom occupations.

Even though we can use this approach to study the LACs dynamics, the losses connected to the initial detection reduce significantly the initial occupation probability for the post-selected atom number [see Fig.~\ref{SM Fig.4}], reducing also the range available for observing the LACs dynamics. However, given any initial preparation, our model allows to study the expected evolution of the probabilities using the extracted loss rates.
Correcting for imaging-induced initial losses, i.e. assuming $P_{N_\text{sel}=3}(0)=1$, we calculate the probability dynamics for $N=0,\,1,\,2,\,3$, and compare it with the evolution observed experimentally. When considering the dynamics of $P_3(t)$, the difference is just a constant rescaling due to imaging-induced losses.
However for lower numbers, the LACs pulse induces non-trivial transfers between the different probabilities depending on the initial preparation. 
Comparing the lossless case to our data, we are able to rescale the latter for each time step to the ideal preparation, as shown in Fig.~3(g).

\section{Simulations of the atomic spread signal}
We numerically simulate our high-intensity fast-imaging protocol ab initio, modeling the three-dimensional (3D) atomic trajectory resulting from photon absorption and stochastic re-emission under pulsed illumination, as well as the process of image formation on the camera sensor. 
In short, a simulated image of a trapped atom is produced by breaking the image creation process into two separate steps. 
First, we simulate the stochastic photon scattering along with the atomic trajectory, keeping track of both the location and emission direction of each photon, as well as the atomic position over the entire illumination time. 
This allows to generate a 3D fluorescence pattern in real space. 
Then, such photon emission pattern is mapped onto the camera sensor, accounting for the finite numerical aperture of our objective, the finite optical transmission and the quantum efficiency of the detector.

To generate the fluorescence photon distribution, we consider both the effect of the atomic motion and the dipole emission pattern.
At $t=0$, the atomic position and momentum are randomly drawn considering a 3D harmonic potential, according to a Boltzmann distribution at the experimental temperature. 
The time of a photon scattering event is obtained by sampling the probability distribution of spontaneous emission, which depends also on the excitation probability for a certain imaging beam intensity and Doppler shift from the finite atomic speed. 
For each time step, the time of the next scattering process is extracted. 
If the time extracted is longer than \SI{10}{\nano\second} $\sim 2/\Gamma_{399}$, the atomic trajectory is updated without scattering a photon by taking into account only the trapping force and gravity. 
If instead a scattering event occurs, the atomic momentum is updated by also considering the momentum kicks from photon absorption and emission.
The maximum time step is chosen as a trade-off between simulation speed and a faithful sampling of the atomic trajectory. 
The imaging beams are pulsed during the whole illumination with a single-pulse duration of \SI{400}{\nano\second}.

The image revealed by the camera sensor depends on the optical propagation of the emitted light throughout the imaging system.
In our simulations, we select photons within the solid angle associated with the objective entrance aperture based on their emission direction, and we apply a stochastic loss process due to finite optical transmission of the objective. 
Subsequently, the 3D position where each photon has been emitted is converted into a two-dimensional position on the imaging plane by considering the point spread function (PSF) of the imaging system. 
The PSF also accounts for our imaging system magnification of about $\times 8$. 
For this, we apply Monte Carlo sampling of a 2D cut of the PSF of each emitted photon in the imaging plane. 
As each photon's PSF depends on the precise point in space where the photon has been emitted, this approach correctly describes the out-of-focus blurring associated with atoms exiting the focal plane of the imaging objective during their stochastic motion. %($\sim$\SI{1}{\micro\meter}). 
Finally, the resulting photon distribution is integrated on square regions with a size of \SI{4.6}{\micro\meter} $\times$ \SI{4.6}{\micro\meter}, corresponding to the camera pixels. 
The final simulated image is generated by further accounting for the finite quantum efficiency of each camera pixel and the camera pixel readout noise.

We exploit imaging simulations to directly check the effect of the photon-recoil induced atomic motion along the optical axis of the objective, in order to determine if this leads to an appreciable additional blurring of the atomic signal. 
We quantify blurring effects by fitting the simulated atom signal with a Gaussian function in different scenarios. 
First, we check that the theoretical PSF of the imaging system is reproduced in simulations for a fixed motionless atom. 
We then simulate the confinement produced by our tweezer traps, and also test the case where a vertical optical lattice is added to strongly increase the axial tweezer confinement, reducing the vertical (out-of-focus) atomic motion. 
For $t_\text{ill} \lesssim20\,\text{\textmu s}$, simulation results for the three cases above are qualitatively similar, with atomic motion leading to fading of the Airy disc structure, but not to any visible increase of the atomic signal size.
We thus conclude that the atomic motion along imaging optical axis does not compromise the imaging resolution for our tweezer traps and illumination times.

\begin{figure}[t]    \centering\includegraphics[width=0.65\columnwidth]{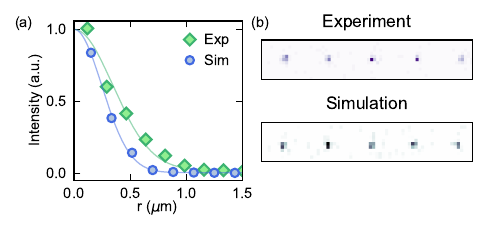}
    \caption{Comparison between experimental and simulated images. (a) Size of the atomic signal for experimental and simulated images. (b) Single shot experimental image and simulation of a 10-sites array. For both the experiment and the simulation we employ an illumination time of $\sim$\,\SI{12}{\us} and a \SI{2.3}{\milli \kelvin} trap depth.}  
    \label{SM Fig.5}
\end{figure}

On the other hand, the finite size of the camera pixels strongly affects our effective imaging resolution: convolving the imaging system PSF with a square window function corresponding to the pixel dimension leads to an increase in the atom signal size, with the typical minima of the Airy disk completely disappearing.
For this reason, even for a diffraction-limited imaging system, we are not able to obtain images that match the optical signal spread. 
This also reduces the impact of atomic motion on the atomic signal, since the blur due to atomic motion is smaller than the effect of pixel convolution. 
By simulating a finer pixel grid, corresponding to a larger magnification factor of the imaging system, we check that for long illumination times, a visible blur eventually leads to visible deviations of the atomic signal with respect to the ideal PSF [see Fig. 4(a)].

Finally, we compare simulated images of an atom array to the experimental ones. We find that the observed spot size in our setup is only marginally larger than the expected diffraction-limited signal, likely due to small residual aberrations in the imaging system [Fig.~\ref{SM Fig.5}(a)]. In case of single shots, comparable photons distributions are recorded for the experimental and the simulated image [Fig.~\ref{SM Fig.5}(b)].

\section{Max-likelihood algorithm for number-resolved detection in dense arrays}

In tightly-spaced tweezer arrays and in lattice gas microscopes, the signal from neighboring atoms starts to overlap, rendering the definition of independent ROIs around individual sites infeasible.
Therefore, straightforward threshold-based site-occupation reconstruction methods fail, and more advanced approaches are required.
Possible refined analysis methods include machine learning-based approaches~\cite{Impertro_2023, Picard_2019} as well as statistical analysis methods. 
Here, we opt for the latter, as it allows us to fully utilize our detailed knowledge of the expected signal distributions. 
Specifically, we implement a Bayesian analysis framework based on the maximum likelihood approach. 
To this end, we start by formulating a set of (global) hypotheses $h$ on the occupation of the tweezer array of $N$ tweezers $\ket{h}=\ket{n_1}\ket{n_2}\cdots\ket{n_N}$, where $n_i$ denotes the occupation of the $i^{th}$ tweezer. 
Based on prior knowledge of the range of expected occupation numbers (e.g. based on the LACs pulse duration in our setup), we can restrict $n_i \in [0,1,2]$ (or $n_i \in [0,1]$). 
From our experimental data, we can also accurately reconstruct the effective atom spread function, which includes diffusion effects in addition to the optical signal spread. 
Further, we assume the scattered photon number for each atom to follow a Poissonian distribution, which allows us to formulate the probability distribution $P_1(n_\text{photon}|h)$ describing the expected number of photons $n_\text{photon}$ impinging on a given pixel (indexed by $x_i$), conditioned on the hypothesis $h$.
Finally, we can characterize the distribution of noise and background counts recorded by the camera, which is directly extracted from a set of experimental images, and therefore we are able to formulate the probability $P_2(n_\text{counts}(x_i)|n_\text{photon}(x_i))$ that a certain number of counts $n_\text{counts}(x_i)$ is recorded on camera pixel $x_i$, conditioned on the number of signal photons $n_\text{photon}(x_i)$ impinging on this pixel.
Based on these two probability distributions, we can estimate the global logarithmic likelihood of a hypothesis by comparing the experimentally measured counts $n^\text{exp}_\text{counts}(x_i)$ for each pixel $x_i$ to the expected distribution for all possible hypotheses:
\begin{align*}
    \mathrm{LogL}(h)&\propto \sum_i\sum_{n_\mathrm{photon}} \log \Big(P_2(n^\mathrm{exp}_\mathrm{counts}(x_i)|n_\mathrm{photon}(x_i))\, 
    \cdot P_1(n_\mathrm{photon}(x_i)|h)\Big)
\\ \mathrm{LogL}&=\mathrm{max}_h\,\mathrm{LogL}(h).
\end{align*}

There are several key advantages to this approach. 
Firstly, and in stark contrast to ROI based methods, it is a global reconstruction scheme and therefore naturally takes 'leakage' of signal from neighboring tweezers into account.
Secondly, the available spatial information on the atomic signal is fully utilized in the reconstruction.
Finally, this approach also incorporates all available knowledge of fluctuations of photon numbers and background counts as well as the camera readout. 
Therefore, it is a suitable approach even when there is (known) spatially varying gain or noise of the camera, or when the readout statistics of the camera itself is more complex (with the most notable example being electron-multiplying CCD-based imaging methods). 

\begin{figure*}[t]
    \centering
    \vspace{0pt}
    \includegraphics[width=1\textwidth]{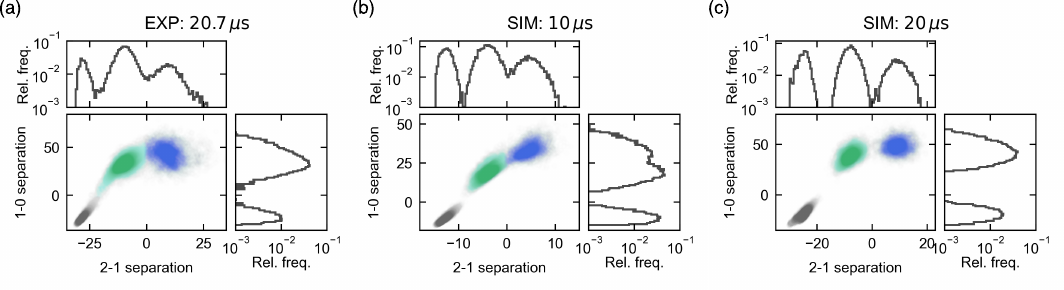}
    \caption{Max-likelihood algorithm for multiply-filled traps. The 2-1 separation is compared to the 1-0 separation (see text), allowing us to identify tweezers to likely be doubly occupied (blue), singly occupied (green) or empty (grey). The relative frequency of the 2-1 separation values (top panel) and 1-0 separation (right panel) is shown as well. (a) Experimental dataset with \SI{20.7}{\micro s} illumination time. (b) Simulated dataset with \SI{10}{\micro s} illumination time. (c) Simulated dataset with \SI{20}{\micro s} illumination time. }
    \label{SM Fig.6}
\end{figure*}

It should be noted that in the limiting case, where the model used in the reconstruction faithfully describes the images, there is a direct correspondence between the posterior distribution of likelihoods and the Fisher-information of the underlying probability distribution~\cite{Hartigan_1983}.
For this reason, in addition to the most likely occupation used in this work, the posterior distribution can be utilized as well, for example to construct weighted correlation functions of site occupations. 
This is particularly helpful when working in a regime where there is not enough information present in the images to unambiguously reconstruct individual occupations.
Finally, this method is neither restricted to lattice systems, nor to the specific imaging scheme employed in this work, and can be directly adapted to other imaging methods including free-space time-of-flight imaging~\cite{klemt_2021}.

In the following, we provide additional information on our capability of number-resolved detection of multiply-filled traps for a tight spacing of $\sim$\,\SI{1.5}{\um}, as shown in the main text. 
For the image analysis, we first calculate the most likely configuration according to our model. 
In order to then evaluate the likelihood separation, for each image and each tweezer individually, we also evaluate the likelihood of the hypothesis where the occupation is set to zero, one, or two (with the remaining tweezers kept at the their most likely occupation, respectively).
As shown in Fig.~\ref{SM Fig.6}, this allows us to define the likelihood separation between the two-atom and one-atom hypothesis (2-1 separation) as well as the zero-atoms to one-atom separation (1-0 separation). 
Note that the experimental dataset shown in Fig~\ref{SM Fig.6} corresponds to the data shown in Fig~4(d) of the main text. For this dataset ($t_\text{ill} = 20.7\, \text{\textmu s}$ and trap depth $\simeq 1.1$\,mK), we observe the presence of small discrepancies between the experimental signal shape and the model used in the reconstruction, in stark contrast to datasets with shorter illumination time (as shown in Fig.~4(b) of the main text). 
We attribute these to percent-level losses during imaging with $t_\text{ill} = 20.7\, \text{\textmu s}$. 
As we have shown very good agreement between simulated and experimental images in the regime where these additional loss mechanisms are negligible (see Fig.~4(c) of the main text), as a reference, we show in simulation that without the additional losses the likelihood distribution between zero, one and two atoms hypotheses, calculated for an illumination time of \SI{20}{\us} [Fig.~\ref{SM Fig.6}(c)], fully separates.
For a shorter illumination time of \SI{10}{\micro s} [Fig.~\ref{SM Fig.6}(b)], where also experimentally losses should become negligible, a very good separation is maintained. 
We therefore expect to be able to increase the fidelity of correctly identifying multiply-filled tweezers in a tightly-spaced array significantly, by either incorporating an additional loss model into the reconstruction or by optimizing the duration of imaging pulses. 
In addition, multiple short-illumination images could be acquired in series (exploiting the fast kinetic mode of electron-multiplying CCD cameras) to leverage time-resolved detection and adaptive maximum likelihood estimation~\cite{Hume_2007,Myerson_2008}.
    
\bibliographystyle{apsrev4-1_ourstyle}
\bibliography{Fast_imaging}